\documentclass[twocolumn,showpacs,floats,floatfix,superscriptaddress,aps,pra]{revtex4}
\usepackage{amsfonts}
\usepackage{amssymb}
\usepackage{amsmath}
\usepackage{graphicx}
\usepackage{bm}

\def\be{\begin{equation}}
\def\ee{\end{equation}}

\begin{document}

\title{Extension of the Morris-Shore transformation to multilevel ladders}
\author{A. A. Rangelov}
\affiliation{Department of Physics, Sofia University, James Bourchier 5 blvd., 1164
Sofia, Bulgaria}
\author{N. V. Vitanov}
\affiliation{Department of Physics, Sofia University, James Bourchier 5 blvd., 1164
Sofia, Bulgaria}
\affiliation{Institute of Solid State Physics, Bulgarian Academy of Sciences,
Tsarigradsko chauss\'{e}e 72, 1784 Sofia, Bulgaria}
\author{B. W. Shore}
\affiliation{Fachbereich Physik der Universit\"{a}t, Erwin-Schr\"{o}dinger-Str., 67653
Kaiserslautern, Germany}
\affiliation{Permanent address: 618 Escondido Cir., Livermore, CA}

\begin{abstract}
We describe situations in which chains of $N$ degenerate quantum energy
levels, coupled by time-dependent external fields, can be replaced by
independent sets of chains, of length $N$, $N-1$, $\ldots ,2$, and sets of
uncoupled single states. The transformation is a generalization of the
two-level Morris-Shore transformation [J.R. Morris and B.W. Shore, Phys.
Rev. A \textbf{27}, 906 (1983)]. We illustrate the procedure with examples
of three-level chains.
\end{abstract}

\maketitle

%====================================

%--------------------------------------------

\section{Introduction}

%--------------------------------------------

The two-level atom has become over the years the basic building block with
which one describes resonant and near-resonant radiative excitation of atoms
and molecules -- or any system that has discrete quantum states \cite%
{All75,Sho90}. %ccccccccc
When the radiation emanates from a laser, and therefore retains temporal
coherence, the relevant dynamics is governed by the time-dependent Schr\"{o}%
dinger equation. Within the usual rotating-wave approximation (RWA) \cite%
{Sho90}, %ccccccccc
the needed mathematics is that of two coupled linear ordinary differential
equations for two complex-valued time-dependent probability amplitudes. The
relative simplicity of these two equations has enabled researchers to find a
variety of analytic solutions, descriptive both of steady radiation
intensity and pulsed excitation by a variety of analytic forms for the
pulses. In essence, one is able to map the physics of the two-level atom,
under suitable conditions, onto the wealth of special functions studied by
19th century mathematicians.

This basic two-state atom has an interesting extension, from a nondegenerate
two-level system (one with just two quantum states -- ground and excited),
to one with a degenerate ground level and a degenerate excited level. This
situation occurs quite commonly for isolated atoms and molecules; these can
be taken to be in states of well defined angular momentum $J$, for which
rotational symmetry produces a degeneracy of $2J+1$ magnetic sublevels. In
RWA the degeneracies can also occur for more general multistate quantum
systems, as we shall note.

For laser-induced transitions between states of angular momentum, one must
consider the several magnetic sublevels, labeled by $M$, that may occur as
possible initial states. Each of them has a possible laser-driven excitation
route into excited magnetic sublevels. If the angular momenta of the ground
level and the excited level are, respectively, $J$ and $J^{\prime }$ (with $%
J=J^{\prime }$ or $J^{\prime }\pm 1$ for electric-dipole transitions), and
if all excited magnetic sublevels are linked with some ground level, then in
general one must consider $2J+1+2J^{\prime }+1$ probability amplitudes,
coupled to one another by radiative interactions.

For general polarization of the laser field selection rules restrict the
change in magnetic quantum number to quantum states whose magnetic quantum
numbers $M$\ and $M^{\prime }$ differ by $\Delta M=-1$, 0 or +1. For an
arbitrary choice of quantization axis, and elliptically polarized light, the
excitation can take place via all of these linkages. However, when the
polarization is more specialized, to linear or circular, then it is possible
to choose a quantization axis such that these coupled equations become a set
of $2J+1$ pairs of independent two-state equations. The choice of
quantization axis, together with the properties of the rotation matrix of
angular momentum states, makes this possible. For such situations the
mathematics is much simpler: one need only find a set of independent
solutions to the nondegenerate two-state systems.

In 1983 Morris and Shore showed \cite{Morris83} %ccccccccc
that this coordinate transformation was a special case of a more general
transformation that could produce, for any two degenerate sets of quantum
states, an equivalent description involving only independent uncoupled pairs
of equations. Specifically, the Morris-Shore (MS) transformation reduces the
coherent quantum dynamics of a coupled degenerate two-level system to a set
of independent nondegenerate two-state systems and a number of uncoupled
(dark) states. It prescribes a simple recipe, which only requires to find
the eigenvalues and the eigenstates of a hermitean matrix, which is a
product of interaction matrices. The eigenstates are the MS states (i.e. the
states representing the independent two-state systems and the dark states),
and the eigenvalues are the MS interactions in the independent MS two-state
systems. The MS\ transformation requires that all initial interactions be
constant, or share the same time dependence, and that all interactions are
resonant, or equally detuned from the upper states, so that each pair of
interactions are on two-photon resonance with the corresponding states; the
latter condition implies that the lower set of states is degenerate in RWA
sense, and so is the upper set of states.

The MS transformation has been used extensively in various excitation
scenarios to handle seemingly complicated linkages. For instance, it has
been used to derive exact analytic solutions that extend known two-state
solutions to degenerate two-state systems \cite{MS-analytic}. %ccccccccc
It has been used to design schemes for complete population transfer between
degenerate states \cite{MS-CPT} %ccccccccc
and for creation of coherent superpositions of states \cite{MS-sup}.
%ccccccccc
The MS transformation has been also a crucial analytic method in creating
recipes for efficient discrete quantum state tomography \cite{MS-tomo}.
%ccccccccc

There are many situations in which one is interested in transitions that
link not just two states but a chain-like sequence of multilevel
excitations. Typically these form an $N$-level ladder linkage pattern,
involving with each link a separate laser field. When such a system has
degeneracy only from angular momentum, and when the pulses all share a
common time-dependent envelope (though not the same carrier frequencies),
and when the polarizations are all linear or circular, then it is possible
to choose the (arbitrary) quantization axis such that the entire excitation
scheme can be reduced to $2J+1$\ sets of $N$\ coupled equations, where $J$\
is the angular momentum of the initially populated level.

It is natural to ask whether the MS transformation of two levels has a
generalization to $N$-level ladders. We here answer this question
affirmatively, with the proviso of certain conditions, and describe the
procedure for finding the transformation. Such a transformation allows one
to use the well-known analytic solutions of the $N$-state ladder \cite%
{Bia77,Ebe77a,Sho81b} %ccccccccc
as an extension of the utilization of analytic two-state solutions.

This paper is organized as follows. In Sec. \ref{Sec-2levels} we review the
two-level MS transformation and set the stage for its extension. In Sec. \ref%
{Sec-quasi2levels} we describe a resonantly coupled multilevel ladder, which
is reducible to the two-level case. Section \ref{Sec-3levels} presents in
detail the most general, non-resonant extension of the MS\ transformation to
three levels, and Sec. \ref{Sec-Nlevels} extends these results to $N$
levels. Finally, Sec. \ref{Sec-conclusions} presents a summary of the
results.

%--------------------------------------------

\section{The two-level Morris-Shore transformation\label{Sec-2levels}}

%--------------------------------------------

The original MS transformation adopts a state ordering wherein the $N_{a}$
sublevels of the $a$ level are placed first, followed by the $N_{b}$
sublevels of the $b$ level. This allows us to view the RWA\ Hamiltonian as a
block matrix,
\begin{equation}
\mathsf{H}(t)=\left[
\begin{array}{ll}
\mathsf{O} & \mathsf{V}(t) \\
\mathsf{V}(t)^{\dagger } & \mathsf{D}%
\end{array}%
\right] .  \label{Ham}
\end{equation}%
Here $\mathsf{O}$ is the $N_{a}$-dimensional square zero matrix, in which
the zero off-diagonal elements reflect the absence of single-photon
couplings between the $a$ states, while the zero diagonal elements show that
the $a$ states have the same energy, taken as the zero of the energy scale.
The matrix $\mathsf{D}$ is a $N_{b}$-dimensional square diagonal matrix,
which can be represented as a constant multiple of the $N_{b}$-dimensional
unit matrix $\mathsf{1}_{N_{b}}$, $\mathsf{D}=\Delta \mathsf{1}_{N_{b}}$.
The absence of off-diagonal elements in $\mathsf{D}$ reflects the absence of
direct couplings between the $b$ states, while the common diagonal elements $%
\Delta $ stand for the common detunings of all the $b$ states: by
definition, $\Delta $ is the difference between the Bohr transition
frequency and the laser carrier frequency. The $N_{a}\times N_{b}$ matrix $%
\mathsf{V}(t)$ comprises the interactions of the $a$ subevels with the $b$
sublevels; these may depend on time, but the time dependence \emph{must be
the same for every element}. As evident from Eq. (\ref{Ham}), it is assumed
that there are no relaxation processes during the interaction.

%%%%%%%%%%%%%%%%%%%%%%%%%%%%%%%%%%%%
\begin{figure}[tb]
\includegraphics[width=50mm]{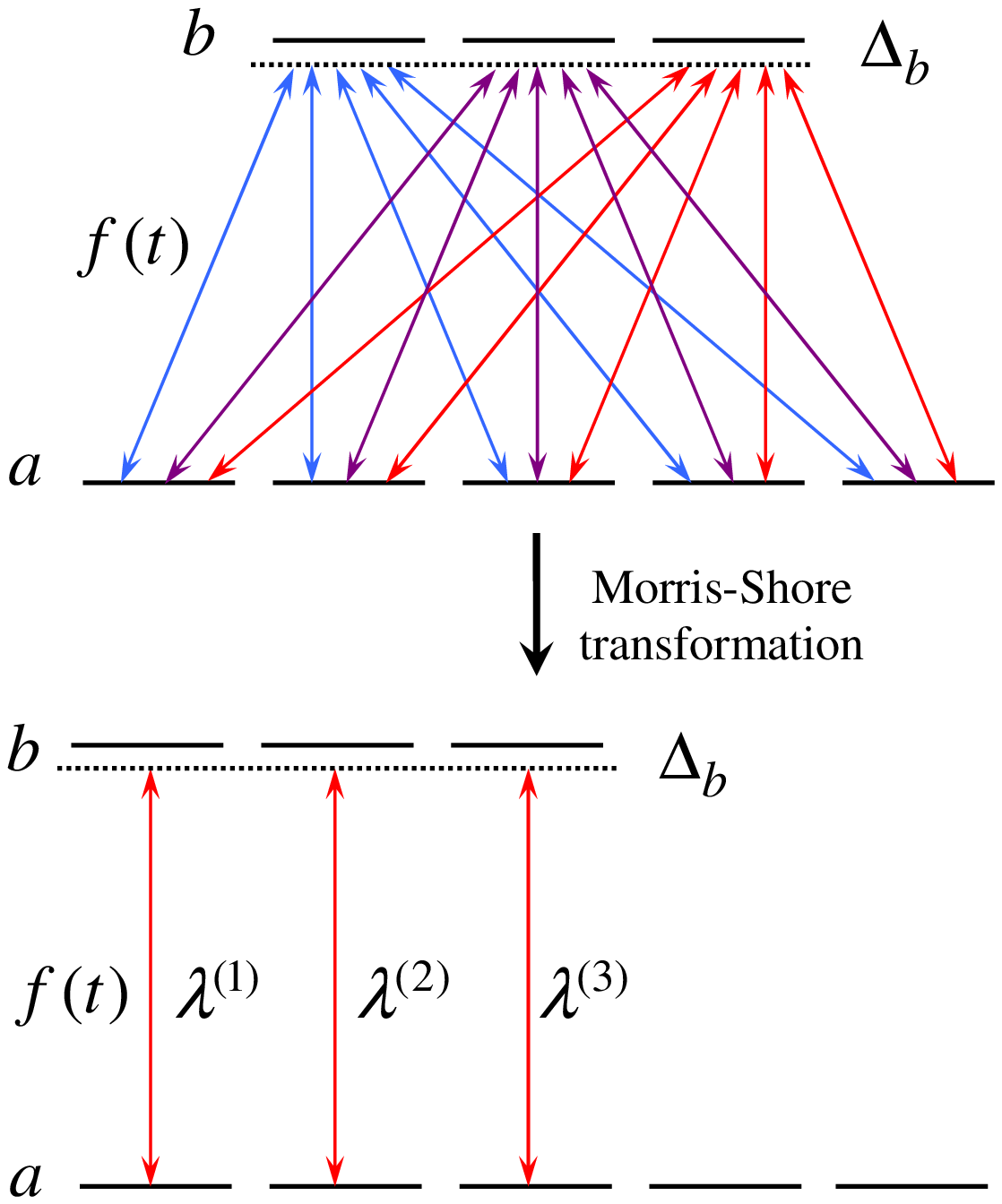}
\caption{(Color online) The essence of the Morris-Shore transformation: a
multistate system consisting of two coupled sets of degenerate levels is
decomposed into a set of uncoupled nondegenerate two-state systems and a set
of uncoupled (dark) states.}
\label{Fig-ms2}
\end{figure}
%%%%%%%%%%%%%%%%%%%%%%%%%%%%%%%%%%

Although we shall discuss this Hamiltonian within the framework of angular
momentum degeneracy, the results are applicable much more generally, to
linkage patterns between nondegenerate states, as will occur when there are
several different laser frequencies, each resonant (or near resonant) with a
particular transition. When such situations are eligible for description by
a multilevel RWA \cite{Sho90}, and when the carrier frequencies are such
that at most two nonzero detunings occur in the RWA Hamiltonian, then the MS
transformation can be used. Specifically, Morris and Shore have shown \cite%
{Morris83} %ccccccccc
that any RWA-degenerate two-level system, in which all couplings share the
same time dependence, can be reduced with a constant unitary transformation
to an equivalent system comprising only independent two-state systems and
uncoupled (dark) states, as shown in Fig. \ref{Fig-ms2}. This time
independent transformation is given by
\begin{equation}
\left\vert \psi _{i}\right\rangle =\sum_{k}S_{ki}^{\ast }\left\vert \varphi
_{k}\right\rangle \quad \Longleftrightarrow \quad \left\vert \varphi
_{k}\right\rangle =\sum_{i}S_{ki}\left\vert \psi _{i}\right\rangle .
\label{transformation}
\end{equation}%
The constant transformation matrix $\mathsf{S}$ can be represented in the
block-matrix form
\begin{equation}
\mathsf{S}=\left[
\begin{array}{cc}
\mathsf{S}_{a} & \mathsf{O} \\
\mathsf{O} & \mathsf{S}_{b}%
\end{array}%
\right] ,  \label{W}
\end{equation}%
where $\mathsf{S}_{a}$ is a unitary $N_{a}$-dimensional square matrix and $%
\mathsf{S}_{b}$ is a unitary $N_{b}$-dimensional square matrix, $\mathsf{S}%
_{a}\mathsf{S}_{a}^{\dagger }=\mathsf{S}_{a}^{\dagger }\mathsf{S}_{a}=%
\mathsf{1}_{N_{a}}$ and $\mathsf{S}_{b}\mathsf{S}_{b}^{\dagger }=\mathsf{S}%
_{b}^{\dagger }\mathsf{S}_{b}=\mathsf{1}_{N_{b}}$. The constant matrices $%
\mathsf{S}_{a}$ and $\mathsf{S}_{b}$ mix only sublevels of a given level: $%
\mathsf{S}_{a}$ mixes the $a$ sublevels and $\mathsf{S}_{b}$ mixes the $b$
sublevels. The transformed MS Hamiltonian has the form (to simplify notation
we here omit explicit display of time dependence)
\begin{equation}
\mathsf{H}^{MS}=\mathsf{S}\mathsf{H}\mathsf{S}^{\dagger }=\left[
\begin{array}{cc}
\mathsf{O} & \mathsf{M} \\
\mathsf{M}^{\dagger } & \mathsf{D}%
\end{array}%
\right] ,  \label{H-original}
\end{equation}%
where
\begin{equation}
\mathsf{M}=\mathsf{S}_{a}\mathsf{VS}_{b}^{\dagger }.  \label{M}
\end{equation}%
The $N_{a}\times N_{b}$ matrix $\mathsf{M}$ may have null rows (if $%
N_{a}>N_{b}$) or null columns (if $N_{a}<N_{b}$), which correspond to dark
states; let us assume that $N_{a}>N_{b}$. The decomposition of $\mathsf{H}$
into a set of independent two-state systems requires that, after removing
the null rows or columns, $\mathsf{M}$ reduces (possibly after an
appropriate relabeling) to a diagonal matrix; let us denote its diagonal
elements by $\lambda ^{(n)}$ ($n=1,2,\ldots ,N_{b}$). It follows from Eq. (%
\ref{M}) that
\begin{subequations}
\label{MM}
\begin{eqnarray}
\mathsf{M}\mathsf{M}^{\dagger } &=&\mathsf{S}_{a}\mathsf{V}\mathsf{V}%
^{\dagger }\mathsf{S}_{a}^{\dagger },  \label{MM+} \\
\mathsf{M}^{\dagger }\mathsf{M} &=&\mathsf{S}_{b}\mathsf{V}^{\dagger }%
\mathsf{VS}_{b}^{\dagger }.  \label{M+M}
\end{eqnarray}%
Hence $\mathsf{S}_{a}$ and $\mathsf{S}_{b}$ are defined by the condition
that they diagonalize $\mathsf{V}\mathsf{V}^{\dagger }$ and $\mathsf{V}%
^{\dagger }\mathsf{V}$, respectively. Because, by assumption, all elements
of $\mathsf{V}$ have the same time dependence $f(t)$, this dependence is
factored out and therefore $\mathsf{S}_{a}$ and $\mathsf{S}_{b}$ are
constant; the eigenvalues, however, are proportional to $f^{2}(t)$ and hence
they depend on time.

It is straightforward to show that the $N_{b}$ eigenvalues of $\mathsf{V}%
^{\dagger }\mathsf{V}$ are all non-negative; according to Eq. (\ref{MM})
they are $\left[ \lambda ^{(n)}\right] ^{2}$. The matrix $\mathsf{V}\mathsf{V%
}^{\dagger }$ has the same eigenvalues and additional $N_{a}-N_{b}$ zero
eigenvalues. The independent two-state systems $\varphi
_{a}^{(n)}\leftrightarrow \varphi _{b}^{(n)}$ ($n=1,2,\ldots ,N_{b}$), each
composed of an $a$ state $\varphi _{a}^{(n)}$ and a $b$ state $\varphi
_{b}^{(n)}$, are driven by the (time-varying) RWA Hamiltonians
\end{subequations}
\begin{equation}
\mathsf{H}^{(n)}=\left[
\begin{array}{cc}
0 & \lambda ^{(n)} \\
\lambda ^{(n)} & \Delta%
\end{array}%
\right] \quad (n=1,2,\ldots ,N_{b}).  \label{Hn}
\end{equation}%
Each of these two-state Hamiltonians has the same detuning $\Delta $; they
differ in the Rabi frequency $2\lambda ^{(n)}$.

%--------------------------------------------

\section{A multilevel Morris-Shore transformation: the quasi-two-level case
\label{Sec-quasi2levels}}

\begin{figure}[tb]
\includegraphics[width=60mm]{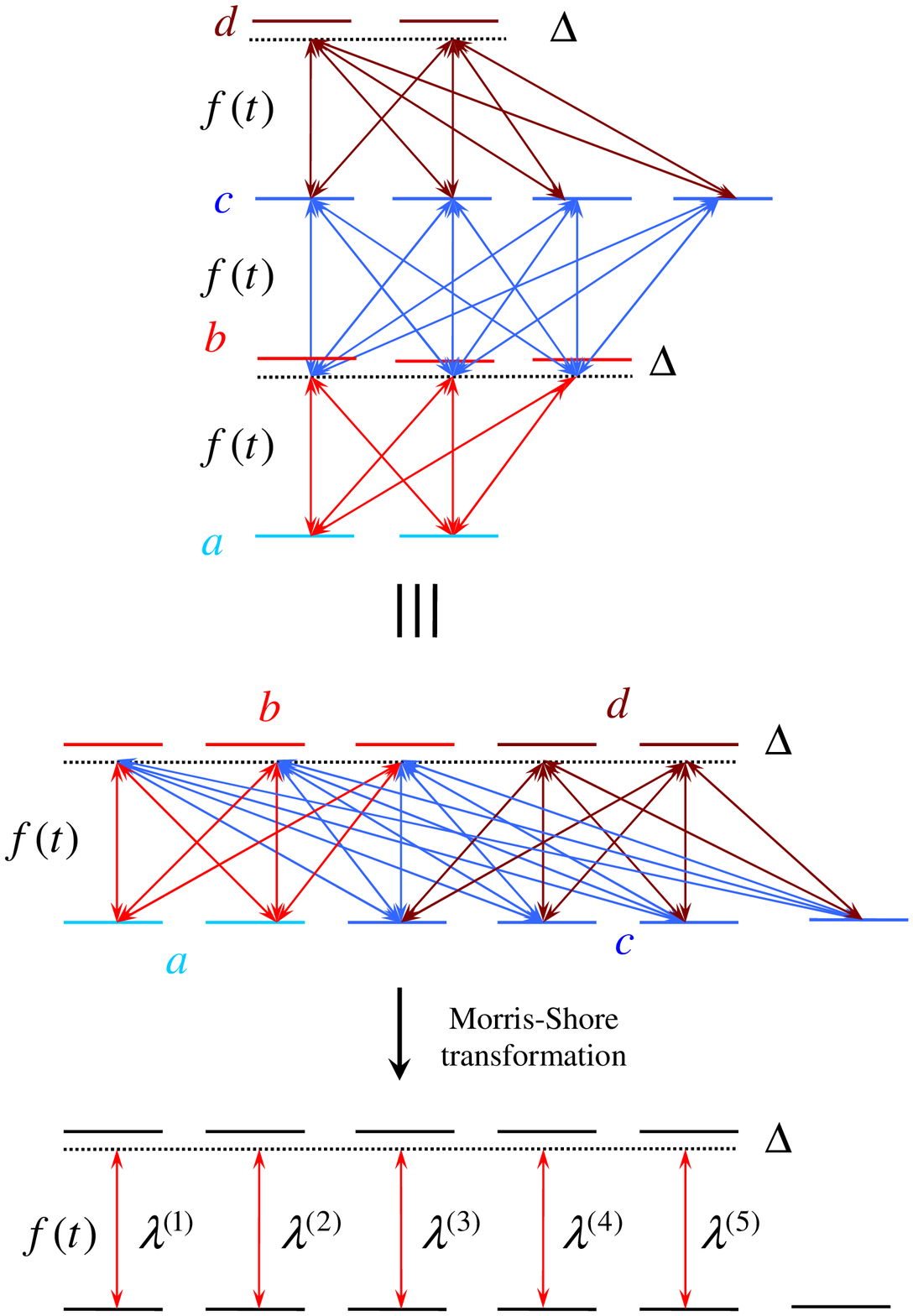}
\caption{(Color online) The reducible multilevel Morris-Shore transformation
in a multistate system consisting of $N$ coupled sets of degenerate levels
when the two-photon resonances $a-c-\cdots $ and $b-d-\cdots $ are fulfilled
(top). All interactions have the same time dependence $f(t)$. We first
formally combine the RWA-degenerate sets $a,c,\ldots $ into one larger
(\textquotedblleft lower\textquotedblright ) set of RWA-degenerate states,
and the sets $b,d,\ldots $ into another larger (\textquotedblleft
upper\textquotedblright ) set of RWA-degenerate states (middle). Then the
new degenerate two-level system is decomposed into a set of independent
nondegenerate two-state systems and a set of uncoupled (dark) states
(bottom).}
\label{Fig-msn}
\end{figure}

The two-level MS transformation is readily extended to multiple degenerate
levels $a-b-c-d-\cdots $ when two conditions are fulfilled: (i) all
couplings share the same time dependence (in particular, all couplings may
be constant), and (ii) the two-photon resonances $a-c-\cdots $ and $%
b-d-\cdots $ are fulfilled (in particular, all fields may be on resonance
with the respective transition frequency). This can be achieved by formally
combining the RWA-degenerate sets $a,c,\ldots $ into one larger set of
RWA-degenerate states, and the sets $b,d,\ldots $ into another larger set of
RWA-degenerate states. Then one can carry out the MS\ factorization on the
new degenerate two-level system, as displayed in Fig. \ref{Fig-msn}. Then
the MS states in the lower set will be superpositions of $a,c,\ldots $
states, whereas the MS states in the upper set will be superpositions of $%
b,d,\ldots $ states.

%--------------------------------------------

%%%%%%%%%%%%%%%%%%%%%%%%%%%%%%%%%%%%
%%%%%%%%%%%%%%%%%%%%%%%%%%%%%%%%%%

When the above conditions (i) or (ii) are not met, then we cannot reduce the
multilevel case to a two-level one. Nevertheless, it may still be possible
to replace the complicated linkages by simple sets of independent ladders.
The next section presents a truly multilevel extension of the MS
transformation that produces this reduction.

%--------------------------------------------

\section{The three-level Morris-Shore transformation\label{Sec-3levels}}

%--------------------------------------------

%--------------------------------------------

\subsection{The RWA Hamiltonian}

%--------------------------------------------

We consider excitation by a set of coherent laser pulses of a multilevel
system for which the generalized RWA is applicable. The excitation dynamics
is governed by the time-dependent Schr\"{o}dinger equation for the coupled
probability amplitudes $C_{n}(t)$. In matrix form it reads
\begin{equation}
i\frac{d}{dt}\mathbf{C}(t)=\mathsf{H}(t)\mathbf{C}(t).
\end{equation}%
The elements of the RWA Hamiltonian matrix $\mathsf{H}(t)$ (in units of $%
\hbar $) are detunings (on the diagonal) and time-dependent Rabi frequencies
times $1/2$. For simplicity we shall, in the following, omit explicit time
arguments.

\begin{figure}[tb]
\includegraphics[width=50mm]{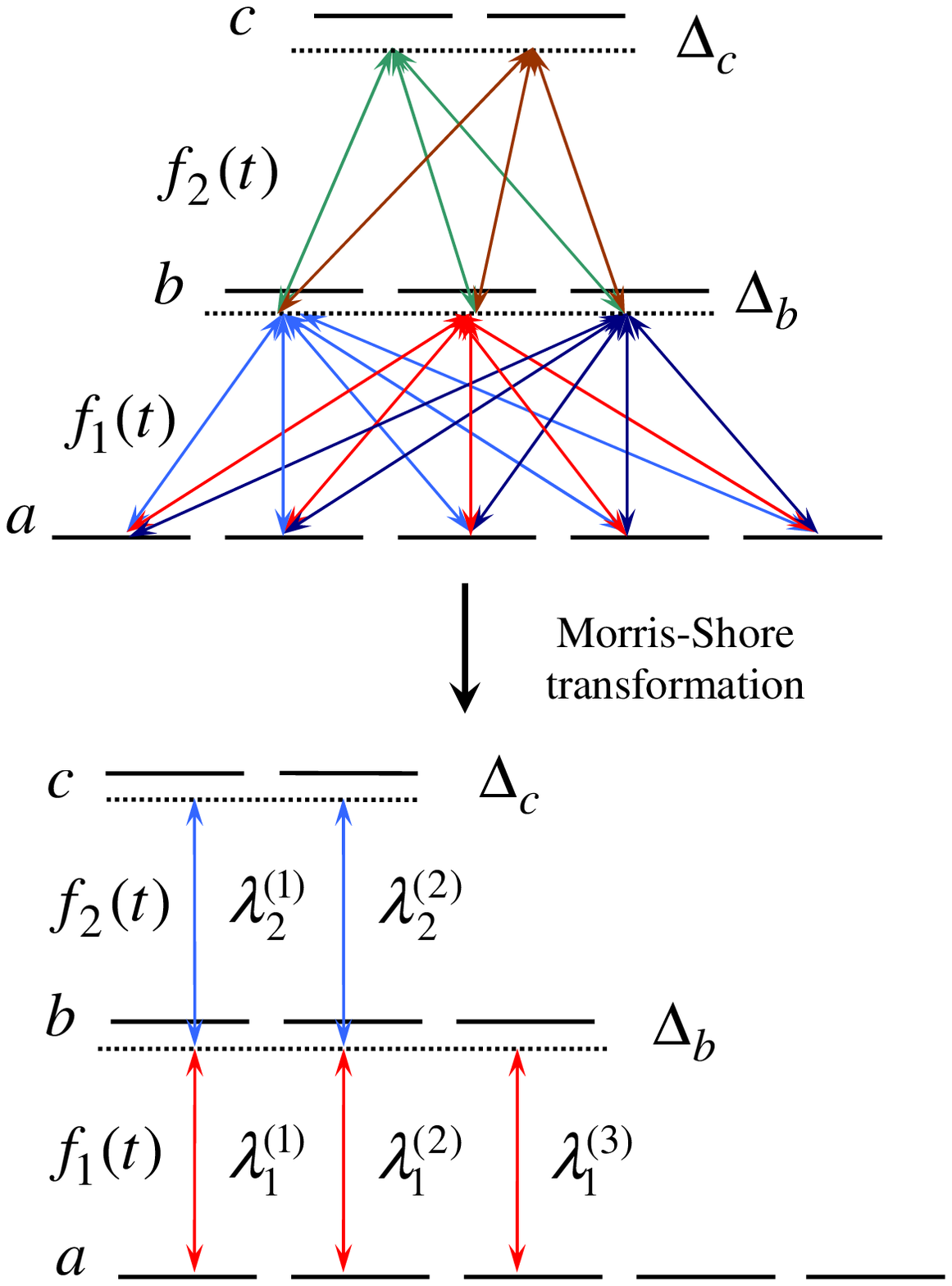}
\caption{(Color online) The Morris-Shore transformation for three degenerate
levels: the MS decomposition produces sets of independent nondegenerate
three-state and two-state systems and a set of uncoupled (dark) states,
provided the commutation condition (\protect\ref{commutator}) is fulfilled.}
\label{Fig-ms3}
\end{figure}

Let us specialize this equation to a three-level system, wherein there are $%
N_{k}$ degenerate sublevels of level $k$, where $k$ runs over indices $a$, $%
b $ and $c$. For definiteness we assume that these degenerate levels form a
ladder, i.e. $E_{a}<E_{b}<E_{c}$. Figure \ref{Fig-ms3} shows a possible
linkage pattern amongst the quantum states: those of level $a$ link only to
those of level $b$, as do those of level $c$; we assume there are no direct
linkages between the $a$ states and the $c$ states. These assumptions allow
us to present the RWA Hamiltonian in the block-matrix form
\begin{equation}
\mathsf{H}=\left[
\begin{array}{ccc}
\mathsf{O} & \mathsf{V}_{1} & \mathsf{O} \\
\mathsf{V}_{1}^{\dag } & \mathsf{D}_{b} & \mathsf{V}_{2} \\
\mathsf{O} & \mathsf{V}_{2}^{\dag } & \mathsf{D}_{c}%
\end{array}%
\right] .  \label{H}
\end{equation}%
Here the matrix $\mathsf{O}$ in the upper left corner is a $N_{a}$%
-dimensional square null matrix, where the null off-diagonal elements
reflect the absence of radiative couplings amongst the $a$ sublevels, while
the null diagonal elements originate with our (conventional) choice of RWA
phases. The null matrices in the upper right and lower left corners indicate
the absence of direct linkages between the $a$ states and the $c$ states.
The square matrices $\mathsf{D}_{b}$ and $\mathsf{D}_{c}$ are scalar
multiples of unit matrices of dimensions $N_{b}$ and $N_{c}$, respectively, $%
\mathsf{D}_{b}=\Delta _{b}\mathsf{1}_{N_{b}}$ and $\mathsf{D}_{c}=\Delta _{c}%
\mathsf{1}_{N_{c}}$. The scalars $\Delta _{b}$ and $\Delta _{c}$ are,
respectively, the usual one- and two-photon detunings associated with the
RWA. Although not shown explicitly, the interactions $\mathsf{V}_{1}$ and $%
\mathsf{V}_{2}$ may depend upon time. However, the elements of each matrix
must share a common time dependence, say $f_{1}(t)$ for $\mathsf{V}_{1}$ and
$f_{2}(t)$ for $\mathsf{V}_{2}$.

%%%%%%%%%%%%%%%%%%%%%%%%%%%%%%%%%%%%
%%%%%%%%%%%%%%%%%%%%%%%%%%%%%%%%%%

%--------------------------------------------------

\subsection{The MS transformation}

%--------------------------------------------------

We wish to transform the original Hamiltonian (\ref{H}) to a form in which
the radiative couplings occur only in single unlinked chains, of length 2 or
3. That is, we seek a transformed MS basis, linked to the original basis by
the transformation (\ref{transformation}), and a corresponding transformed
MS Hamiltonian, which must appear in the block form
\begin{equation}
\mathsf{H}^{MS}=\mathsf{SHS}^{\dag }=\left[
\begin{array}{ccc}
\mathsf{O} & \mathsf{M}_{1} & \mathsf{O} \\
\mathsf{M}_{1}^{\dag } & \mathsf{D}_{b} & \mathsf{M}_{2} \\
\mathsf{O} & \mathsf{M}_{2}^{\dag } & \mathsf{D}_{c}%
\end{array}%
\right] ,  \label{H1}
\end{equation}%
where $\mathsf{M}_{1}$ and $\mathsf{M}_{2}$ are diagonal matrices
supplemented by null columns or rows.

The transformation must only combine sublevels within a given level.
Therefore it must have the form
\begin{equation}
\mathsf{S}=\left[
\begin{array}{ccc}
\mathsf{S}_{a} & \mathsf{O} & \mathsf{O} \\
\mathsf{O} & \mathsf{S}_{b} & \mathsf{O} \\
\mathsf{O} & \mathsf{O} & \mathsf{S}_{c}%
\end{array}%
\right] ,  \label{R}
\end{equation}%
where $\mathsf{S}_{a}$, $\mathsf{S}_{b}$ and $\mathsf{S}_{c}$ are constant
square unitary matrices of dimensions $N_{a}$, $N_{b}$, and $N_{c}$,
respectively With this transformation the block elements of the transformed
Hamiltonian (\ref{H1}) read
\begin{subequations}
\label{PQ}
\begin{eqnarray}
\mathsf{M}_{1} &=&\mathsf{S}_{a}\mathsf{V}_{1}\mathsf{S}_{b}^{\dag },
\label{P} \\
\mathsf{M}_{2} &=&\mathsf{S}_{b}\mathsf{V}_{2}\mathsf{S}_{c}^{\dag }.
\label{Q}
\end{eqnarray}%
The matrices $\mathsf{M}_{1}$ and $\mathsf{M}_{2}$ may have null rows or
columns; these correspond to dark states. The desired decomposition of $%
\mathsf{H}$ into a set of independent two- or three-state systems requires
that, after removing the null rows or columns, $\mathsf{M}_{1}$ and $\mathsf{%
M}_{2}$ reduce (possibly after an appropriate relabeling) to diagonal
matrices. It follows from Eqs. (\ref{PQ}) that the following matrices are
diagonal:
\end{subequations}
\begin{subequations}
\begin{eqnarray}
\mathsf{S}_{a}\mathsf{V}_{1}\mathsf{V}_{1}^{\dag }\mathsf{S}_{a}^{\dag } &=&{%
\mathsf{M}_{1}\mathsf{M}_{1}}^{\dag }=\text{diag},  \label{PP+} \\
\mathsf{S}_{b}\mathsf{V}_{1}^{\dag }\mathsf{V}_{1}\mathsf{S}_{b}^{\dag } &=&%
\mathsf{M}_{1}^{\dag }{\mathsf{M}_{1}}=\text{diag},  \label{P+P} \\
\mathsf{S}_{b}\mathsf{V}_{2}\mathsf{V}_{2}^{\dag }\mathsf{S}_{b}^{\dag } &=&{%
\mathsf{M}_{2}\mathsf{M}_{2}}^{\dag }=\text{diag},  \label{QQ+} \\
\mathsf{S}_{c}\mathsf{V}_{2}^{\dag }\mathsf{V}_{2}\mathsf{S}_{c}^{\dag } &=&%
\mathsf{M}_{2}^{\dag }\mathsf{M}_{2}=\text{diag}\mathbf{.}  \label{Q+Q}
\end{eqnarray}%
Hence $\mathsf{S}_{a}$ and $\mathsf{S}_{c}$ are defined by the condition
that they diagonalize $\mathsf{V}_{1}\mathsf{V}_{1}^{\dag }$ and $\mathsf{V}%
_{2}^{\dag }\mathsf{V}_{2}$, respectively. The matrix $\mathsf{S}_{b}$ must,
by definition, diagonalize both matrices $\mathsf{W}_{1}$ and $\mathsf{W}%
_{2} $, where

\end{subequations}
\begin{equation}
\mathsf{W}_{1}=\mathsf{V}_{1}^{\dag }\mathsf{V}_{1},\quad \mathsf{W}_{2}=%
\mathsf{V}_{2}\mathsf{V}_{2}^{\dag }.  \label{W1,W2}
\end{equation}%
This can only occur if these two products commute,
\begin{equation}
\left[ \mathsf{W}_{1},\mathsf{W}_{2}\right] =\mathsf{O}.  \label{commutator}
\end{equation}%
Hence ${\mathsf{W}_{1}}$ and $\mathsf{W}_{2}$ must have the same set of
eigenvectors. This set, when normalized, forms the transformation matrix $%
\mathsf{S}_{b}$ for the $b$-state manifold. We shall assume hereafter that
Eq. (\ref{commutator}) is satisfied; we will discuss the implications of
this assumption in Sec. \ref{Sec-commutation}.

It is easy to show that the eigenvalues of $\mathsf{V}_{1}\mathsf{V}%
_{1}^{\dag }$ and $\mathsf{V}_{2}^{\dag }\mathsf{V}_{2}$ are all
non-negative, and hence they can be written as squares of real numbers, $%
\left[ \lambda _{1}^{(n)}\right] ^{2}$ and $\left[ \lambda _{2}^{(n)}\right]
^{2}$, respectively. The matrices $\mathsf{W}_{1}$ and $\mathsf{W}_{2}$ have
the same eigenvalues, except for additional (or missing) zero eigenvalues.

In the MS basis, the description of the dynamics comprises sets of
independent ladders, of length no greater than $N=3$. The three-state
systems, expressing the linkages $a\leftrightarrow b\leftrightarrow c$, are
governed by Hamiltonian matrices of the form
\begin{equation}
\mathsf{H}_{abc}^{(n)}=\left[
\begin{array}{ccc}
0 & \lambda _{1}^{(n)} & 0 \\
\lambda _{1}^{(n)} & \Delta _{b} & \lambda _{2}^{(n)} \\
0 & \lambda _{2}^{(n)} & \Delta _{c}%
\end{array}%
\right] .  \label{H3}
\end{equation}%
Two-state systems $a\leftrightarrow b$, if present, are governed by the
Hamiltonians
\begin{equation}
\mathsf{H}_{ab}^{(n)}=\left[
\begin{array}{cc}
0 & \lambda _{1}^{(n)} \\
\lambda _{1}^{(n)} & \Delta _{b}%
\end{array}%
\right] ,  \label{H2}
\end{equation}%
while two-state linkages $b\leftrightarrow c$ are governed by the
Hamiltonians
\begin{equation}
\mathsf{H}_{bc}^{(n)}=\left[
\begin{array}{cc}
\Delta _{b} & \lambda _{2}^{(n)} \\
\lambda _{2}^{(n)} & \Delta _{c}%
\end{array}%
\right] .
\end{equation}%
Finally, there may be single unlinked states, in any of the three levels;
these can be regarded as being governed by one-dimensional matrices
(scalars) $\mathsf{H}_{a}^{(n)}=0$ or $\mathsf{H}_{b}^{(n)}=\Delta _{b}$ or $%
\mathsf{H}_{c}^{(n)}=\Delta _{c}$.

In general, if the number of states in each initial manifold is different,
we denote the minimum and maximum degeneracies by $N_{\min }\equiv \min
\left\{ N_{a},N_{b},N_{c}\right\} $, $N_{\max }\equiv \max \left\{
N_{a},N_{b},N_{c}\right\} $, and the intermediate number by $N_{\text{mid}}$%
. We can then identify the following possibilities:

\begin{itemize}
\item if $N_{\min }=N_{b}$ then in the MS basis there will be $N_{b}$
three-state systems, $N_{a}-N_{b}$ dark states in the $a$ set of states, and
$N_{c}-N_{b}$ dark states in the $c$ set;

\item if $N_{\min }=N_{a}$ or $N_{c}$ then in the MS basis there will be $%
N_{\min }$ three-state systems, $N_{\text{mid}}-N_{\min }$ two-state systems
composed of states of the sets with $N_{\text{mid}}$ and $N_{\max }$, and $%
N_{\max }-N_{\text{mid}}$ dark states composed of states of the set with $%
N_{\max }$.
\end{itemize}

Figure \ref{Fig-ms3} shows an example in which the MS transformation reduces
a general linkage pattern involving 10 states to a pair of dark states, a
single two-state linkage, and a pair of three-state linkages. The time
dependences $f_{1}(t)$ and $f_{2}(t)$ of the lower and upper transitions are
arbitrary. In particular, the $f_{2}(t)$ interaction may precede the $%
f_{1}(t)$ interaction, as is the case of the STIRAP process \cite{STIRAP}.

%-------------------------------------------------------------------

\subsection{Special case: single intermediate state}

%-------------------------------------------------------------------

\begin{figure}[t]
\includegraphics[width=55mm]{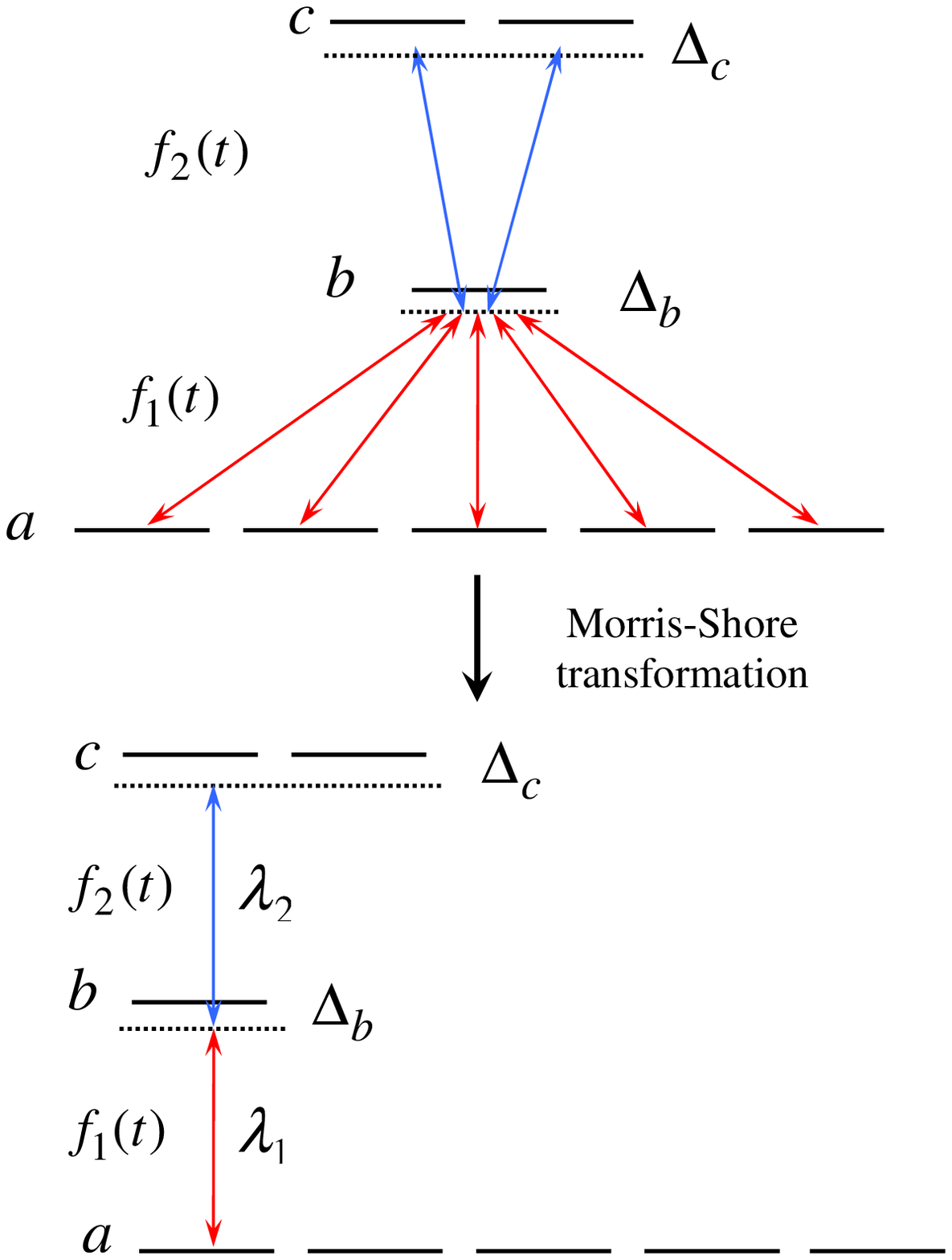}
\caption{(Color online) The Morris-Shore transformation for a single
nondegenerate intermediate state: the MS decomposition produces a
nondegenerate three-state system and two sets of uncoupled (dark) states in
the $a$ and $c$ sets of states.}
\label{Fig-n1m}
\end{figure}

The commutation condition (\ref{commutator}) is fulfilled automatically in
the special case of a single, nondegenerate intermediate state, $N_{b}=1$,
because then the matrices $\mathsf{W}_{1}^{\dag }$ and $\mathsf{W}_{2}$
reduce to scalars. Then, regardless of the degeneracies $N_{a}$ and $N_{c}$
of states $a$ and $c$, the three-level MS transformation always produces a
nondegenerate three-state system comprising a bright state from the $a$
level, a bright state from the $c$ level, and the intermediate state. In
addition, there will be $N_{a}-1$ uncoupled states in the $a$ level and $%
N_{c}-1$ uncoupled states in the $c$ level. Hence for a single intermediate
state the MS\ transformation is \emph{always} possible. Figure \ref{Fig-n1m}
depicts an example of such a linkage pattern and the result of a MS
transformation.

%%%%%%%%%%%%%%%%%%%%%%%%%%%%%%%%%%%%
%%%%%%%%%%%%%%%%%%%%%%%%%%%%%%%%%%

%-------------------------------------------------------------------

\subsection{Consequences of the interaction commutation\label%
{Sec-commutation}}

%-------------------------------------------------------------------

We now turn to the implications of the commutation relation (\ref{commutator}%
). This condition limits the generality of the MS transformation for three
sets of degenerate states. We here pose the question: given the interaction $%
\mathsf{V}_1$, what is the most general form of the interaction $\mathsf{V}%
_2 $, for which the three-level MS transformation applies?

%-------------------------------------------------------------------

\subsubsection{The Frobenius problem}

%-------------------------------------------------------------------

The required commutation relation (\ref{commutator}) is equivalent to
solving a matrix equation of the form%
\begin{equation}
\mathsf{AX}=\mathsf{XA}  \label{Frobenius}
\end{equation}%
for the matrix $\mathsf{X}$, given that $\mathsf{A}$ and $\mathsf{X}$ are
both square Hermitian matrices of the same dimension $N_{b}$. This is known
as the Frobenius problem \cite{Gantmacher}. Because $\mathsf{A}$ and $%
\mathsf{X}$ commute, they have the same set of eigenvectors $\gamma _{n}$,
\begin{subequations}
\begin{eqnarray}
\mathsf{A}\gamma _{n} &=&\alpha _{n}\gamma _{n}, \\
\mathsf{X}\gamma _{n} &=&\xi _{n}\gamma _{n},
\end{eqnarray}%
and they are diagonalized by the same unitary matrix $\mathsf{G}$, composed
of these eigenvectors,
\end{subequations}
\begin{subequations}
\begin{eqnarray}
\mathsf{G}{\mathsf{AG}}^{\dag } &=&\mathsf{A}_{0}=\text{diag}\left\{ \alpha
_{1},\alpha _{2},\ldots ,\alpha _{N_{b}}\right\} ,  \label{GFG} \\
\mathsf{G}{\mathsf{XG}}^{\dag } &=&\mathsf{X}_{0}=\text{diag}\left\{ \xi
_{1},\xi _{2},\ldots ,\xi _{N_{b}}\right\} .  \label{GXG}
\end{eqnarray}%
Hence,
\end{subequations}
\begin{equation}
\mathsf{X}={\mathsf{G}}^{\dag }{\mathsf{X}}_{0}{\mathsf{G}}.  \label{X}
\end{equation}%
We can view these results as follows. Given any Hermitian matrix $\mathsf{A}$%
, we can find the transformation matrix $\mathsf{G}$ which diagonalizes it.
Then the most general form of the matrix $\mathsf{X}$ is the construction of
Eq. (\ref{X}), where the real diagonal matrix $\mathsf{X}_{0}$ is arbitrary.
Therefore, if $\mathsf{A}$ and $\mathsf{X}$ are $N_{b}$-dimensional then the
matrix $\mathsf{X}$ is parametrized by $N_{b}$ parameters: the diagonal
elements of $\mathsf{X}_{0}$.

Alternatively, we can express the matrix $\mathsf{X}$ as a power series in $%
\mathsf{A}$. The Cayley-Hamilton theorem \cite{Gantmacher} implies that only
$N_{b}-1$ of these powers, e.g. $\mathsf{A}^{0}$, $\mathsf{A}^{1}$, $\ldots $%
, $\mathsf{A}^{N_{b}-1}$, are linearly independent. Then the expansion reads
\begin{equation}
\mathsf{X}=\sum_{n=0}^{N_{b}-1}x_{n}\mathsf{A}^{n},  \label{X=polynomial}
\end{equation}%
where the $N_{b}$ coefficients $x_{n}$ are arbitrary. Because $\mathsf{X}$
is Hermitian, these numbers must be real.

Hence either of the two alternative solutions of the Frobenius problem (\ref%
{Frobenius}) -- either Eq. (\ref{X}) or Eq. (\ref{X=polynomial}) -- involve $%
N_{b}$ arbitrary real parameters.

%-------------------------------------------------------------------

\subsubsection{Implications for linkages}

%-------------------------------------------------------------------

We now apply these results to the MS transformation. We know that any given
interaction $\mathsf{V}_{1}$ determines the transformation matrix $\mathsf{S}%
_{b}$ through Eq. (\ref{P+P}). It follows that the interaction $\mathsf{V}%
_{2}$ must satisfy%
\begin{equation}
\mathsf{W}_{2}={\mathsf{S}}_{b}^{\dag }{\mathsf{W}}_{0}{\mathsf{S}}_{b},
\end{equation}%
with $\mathsf{W}_{2}=\mathsf{V}_{2}\mathsf{V}_{2}^{\dag }$, where $\mathsf{W}%
_{0}$ is an arbitrary $N_{b}$-dimensional real diagonal matrix. Equation (%
\ref{X=polynomial}) implies that the most general representation of $\mathsf{%
W}_{2}$, for which the commutation relation (\ref{commutator}) is satisfied
and hence there exists MS transformation, has the form
\begin{equation}
\mathsf{W}_{2}=\sum_{n=0}^{N_{b}-1}w_{n}\mathsf{W}_{1}^{n}.
\end{equation}%
where the $N_{b}$ arbitrary real coefficients $w_{n}$ determine the degrees
of freedom for $\mathsf{W}_{2}$.

%---------------------------------------------------------------

\subsection{Example: $J=3/2\leftrightarrow J=1/2\leftrightarrow J=1/2$ ladder%
}

%---------------------------------------------------------------

\subsubsection{The system and the couplings}

We here illustrate the rather formal results with a specific example. We
consider a three-level ladder whose degeneracy stems from angular momentum.
Specifically we consider the sequence $J=3/2\leftrightarrow
J=1/2\leftrightarrow J=1/2$; hence the magnetic sublevels form a degenerate
three-level system with $N_{a}=4$, $N_{b}=N_{c}=2$. Taking into account the
Clebsch-Gordan coefficients \cite{Zar88} %cccccccc
we find
\begin{subequations}
\begin{eqnarray}
\mathsf{V}_{1}(t) &=&\frac{f_{1}(t)}{\sqrt{6}}\left[
\begin{array}{cc}
r_{1}\sqrt{3} & 0 \\
-p_{1}\sqrt{2} & r_{1} \\
l_{1} & -p_{1}\sqrt{2} \\
0 & l_{1}\sqrt{3}%
\end{array}%
\right] , \\
\mathsf{V}_{2}(t) &=&\frac{f_{2}(t)}{\sqrt{3}}\left[
\begin{array}{cc}
-p_{2} & -r_{2}\sqrt{2} \\
l_{2}\sqrt{2} & p_{2}%
\end{array}%
\right] ,
\end{eqnarray}
where $f_{1}(t)$ anf $f_{2}(t)$ define the (generally different) time
envelopes of the pulsed interactions in the lower and upper transitions,
respectively; $r_{n},p_{n},l_{n}$ are related to the amplitudes (with the
respective phases) of the right-circular ($\sigma ^{+}$), linear ($\pi $),
and left-circular ($\sigma ^{-}$) polarizations for the lower ($n=1$) or
upper ($n=2$) transition.

\subsubsection{The MS states}

The MS states in the $a$ manifold are defined as the eigenstates of the
matrix $\mathsf{V}_{1}\mathsf{V}_{1}^{\dag }$ [see Eq. (\ref{PP+})],%
\begin{widetext}
\begin{equation}
\mathsf{V}_{1}\mathsf{V}_{1}^{\dag }=\frac{f_{1}^{2}(t)}{6}\left[
\begin{array}{cccc}
3\left\vert r_{1}\right\vert ^{2} & -\sqrt{6}p_{1}^{\ast }r_{1} & \sqrt{3}%
l_{1}^{\ast }r_{1} & 0 \\
-\sqrt{6}p_{1}r_{1}^{\ast } & 2\left\vert p_{1}\right\vert ^{2}+\left\vert
r_{1}\right\vert ^{2} & -\sqrt{2}\left( p_{1}l_{1}^{\ast }+p_{1}^{\ast
}r_{1}\right)  & \sqrt{3}l_{1}^{\ast }r_{1} \\
\sqrt{3}l_{1}r_{1}^{\ast } & -\sqrt{2}\left( p_{1}^{\ast
}l_{1}+p_{1}r_{1}^{\ast }\right)  & \left\vert l_{1}\right\vert
^{2}+2\left\vert p_{1}\right\vert ^{2} & -\sqrt{6}p_{1}l_{1}^{\ast } \\
0 & l_{1}r_{1}^{\ast }\sqrt{3} & -\sqrt{6}p_{1}^{\ast }l_{1} & 3\left\vert
l_{1}\right\vert ^{2}%
\end{array}%
\right] .
\end{equation}%
\end{widetext}Two of these eigenstates are dark, with zero eigenvalues,
whereas the other two are bright, with eigenvalues $\left( \lambda
_{1}^{(1)}\right) ^{2}$ and $\left( \lambda _{1}^{(2)}\right) ^{2}$; the
explicit forms of these eigenstates are too cumbersome to be presented here.

The MS states in the $c$ manifold are defined as the eigenstates of the
matrix $\mathsf{V}_{2}^{\dag }\mathsf{V}_{2}$ [see Eq. (\ref{Q+Q})],
\end{subequations}
\begin{equation}
\mathsf{V}_{2}^{\dag }\mathsf{V}_{2}=\frac{f_{2}^{2}(t)}{3}\left[
\begin{array}{cc}
\left\vert p_{2}\right\vert ^{2}-2r_{2}^{\ast }l_{2} & 2\sqrt{2}i\text{Im}%
\left( p_{2}^{\ast }r_{2}\right) \\
2\sqrt{2}i\text{Im}\left( p_{2}^{\ast }l_{2}\right) & \left\vert
p_{2}\right\vert ^{2}-2r_{2}l_{2}^{\ast }%
\end{array}%
\right] .
\end{equation}%
They have eigenvalues $\left( \lambda _{2}^{(1)}\right) ^{2}$ and $\left(
\lambda _{2}^{(2)}\right) ^{2}$. Explicitly, the $\lambda $'s are given by
\begin{subequations}
\label{MS couplings}
\begin{eqnarray}
\lambda _{1}^{(1,2)} &=&\frac{f_{1}(t)}{\sqrt{6}}\eta _{1}\left[ 2\left(
1+\xi _{1}\right) \right.  \notag \\
&&\mp \left. \sqrt{\varepsilon _{1}^{2}+2\xi _{1}\left( 1+\sqrt{%
1-\varepsilon _{1}^{2}}\cos \alpha _{1}\right) }\right] ^{1/2}, \\
\lambda _{2}^{(1,2)} &=&\frac{f_{2}(t)}{\sqrt{3}}\eta _{2}\left[ 1+\xi
_{2}\right.  \notag \\
&&\mp \left. \sqrt{\varepsilon _{2}^{2}+2\xi _{2}\left( 1+\sqrt{%
1-\varepsilon _{2}^{2}}\cos \alpha _{2}\right) }\right] ^{1/2},
\end{eqnarray}%
with $(n=1,2)$
\end{subequations}
\begin{subequations}
\begin{eqnarray}
\varepsilon _{n} &=&\frac{\left\vert l_{n}\right\vert ^{2}-\left\vert
r_{n}\right\vert ^{2}}{\eta _{n}^{2}}, \\
\xi _{n} &=&\frac{\left\vert p_{n}\right\vert ^{2}}{\eta _{n}^{2}}, \\
\eta _{n} &=&\sqrt{\left\vert l_{n}\right\vert ^{2}+\left\vert
r_{n}\right\vert ^{2}}, \\
\alpha _{n} &=&\arg l_{n}^{\ast }r_{n}^{\ast }p_{n}^{2}.
\end{eqnarray}

The MS states in the intermediate level $b$ are the common eigenstates of
the matrices $\mathsf{W}_{1}=\mathsf{V}_{1}^{\dag }\mathsf{V}_{1}$ and $%
\mathsf{W}_{2}=\mathsf{V}_{2}\mathsf{V}_{2}^{\dag }$ [see Eqs. (\ref{P+P})
and (\ref{QQ+})], i.e.,
\end{subequations}
\begin{subequations}
\begin{equation}
\mathsf{W}_{1}=\left[
\begin{array}{cc}
3\left\vert r_{1}\right\vert ^{2}+2\left\vert p_{1}\right\vert
^{2}+\left\vert l_{1}\right\vert ^{2} & -\sqrt{2}\left( p_{1}^{\ast
}r_{1}+p_{1}l_{1}^{\ast }\right) \\
-\sqrt{2}\left( p_{1}r_{1}^{\ast }+p_{1}^{\ast }l_{1}\right) & \left\vert
r_{1}\right\vert ^{2}+2\left\vert p_{1}\right\vert ^{2}+3\left\vert
l_{1}\right\vert ^{2}%
\end{array}%
\right] \frac{f_{1}^{2}(t)}{6},  \label{W1}
\end{equation}%
\begin{equation}
\mathsf{W}_{2}=\left[
\begin{array}{cc}
\left\vert p_{2}\right\vert ^{2}+2\left\vert r_{2}\right\vert ^{2} & -\sqrt{2%
}\left( p_{2}l_{2}^{\ast }+p_{2}^{\ast }r_{2}\right) \\
-\sqrt{2}\left( p_{2}^{\ast }l_{2}+p_{2}r_{2}^{\ast }\right) & \left\vert
p_{2}\right\vert ^{2}+2\left\vert l_{2}\right\vert ^{2}%
\end{array}%
\right] \frac{f_{2}^{2}(t)}{3}.  \label{W2}
\end{equation}%
If $\mathsf{W}_{1}$ and $\mathsf{W}_{2}$ do not commute, then the
eigenstates of $\mathsf{W}_{1}$ will differ from the eigenstates of $\mathsf{%
W}_{2}$ and there will be no MS factorization. In other words, the two-state
MS transformation, when applied to the lower transition $a-b$, will produce
MS states in the $b$ level (defined as the eigenstates of $\mathsf{W}_{1}$),
which will differ from the MS states in this same $b$ level produced by
two-state MS transformation in the upper transition $b-c$ (defined as the
eigenstates of $\mathsf{W}_{2}$). Three-state MS transformation will only
occur if these two sets of $b$ states are the same, a necessary and
sufficient condition for which is the commutation of $\mathsf{W}_{1}$ and $%
\mathsf{W}_{2}$.

\subsubsection{Commutation implications}

The commutation relation (\ref{commutator}) leads to the equations
\end{subequations}
\begin{subequations}
\label{conditions.example}
\begin{gather}
\text{Im}\left[ \left( l_{1}^{\ast }p_{1}+r_{1}p_{1}^{\ast }\right) \left(
l_{2}p_{2}^{\ast }+p_{2}r_{2}^{\ast }\right) \right] =0,  \label{cond.1} \\
\left( l_{1}p_{1}^{\ast }+r_{1}^{\ast }p_{1}\right) (\left\vert
r_{2}\right\vert ^{2}-\left\vert l_{2}\right\vert ^{2})=(\left\vert
r_{1}\right\vert ^{2}-\left\vert l_{1}\right\vert ^{2})\left(
l_{2}p_{2}^{\ast }+p_{2}r_{2}^{\ast }\right) .  \label{cond.2}
\end{gather}%
Obviously, if all interactions are real, the first condition (\ref{cond.1})
is satisfied automatically.

In the general case of complex interactions, one can solve this system of
equations, for example, by considering two cases: when $\left\vert
r_{1}\right\vert \neq \left\vert l_{1}\right\vert $ and $\left\vert
r_{1}\right\vert =\left\vert l_{1}\right\vert $.

(i) For $\left\vert r_{1}\right\vert \neq \left\vert l_{1}\right\vert $, it
is readily seen, by replacing the term $\left( l_{2}p_{2}^{\ast
}+p_{2}r_{2}^{\ast }\right) $ from Eq. (\ref{cond.2}) into Eq. (\ref{cond.1}%
), that Eq. (\ref{cond.1}) is satisfied identically; hence condition (\ref%
{commutator}) requires only one condition to be fulfilled: Eq. (\ref{cond.2}%
). The latter condition can be solved, for example, for $p_{2}$,
\end{subequations}
\begin{equation}
p_{2}=\frac{l_{2}\left( l_{1}^{\ast }p_{1}+r_{1}p_{1}^{\ast }\right)
-r_{2}\left( l_{1}p_{1}^{\ast }+r_{1}^{\ast }p_{1}\right) }{\left\vert
l_{1}\right\vert ^{2}-\left\vert r_{1}\right\vert ^{2}}.  \label{cond.P}
\end{equation}%
Condition (\ref{cond.P}) restricts the amplitude of the linearly polarized
field for the upper transition to be a function of the arbitrary amplitudes
of the other fields. Because $p_{2}$ is complex-valued, condition (\ref%
{cond.P}) represents, in fact, \emph{two} conditions: for the modulus and
the phase of $p_{2}$.

(ii) For $\left\vert r_{1}\right\vert =\left\vert l_{1}\right\vert $, there
are obviously \emph{two} solutions. The \emph{first} of these is $\left\vert
r_{2}\right\vert =\left\vert l_{2}\right\vert $; then Eq. (\ref{cond.1}) is
also required because it is not satisfied automatically. The \emph{second}
solution is $l_{1}p_{1}^{\ast }+r_{1}^{\ast }p_{1}=0$; it fixes one of the
phases of the lower-transition fields (e.g., $2\arg p_{1}=\pi -\arg
l_{1}+\arg r_{1}$).

\begin{figure}[tb]
\includegraphics[width=47mm]{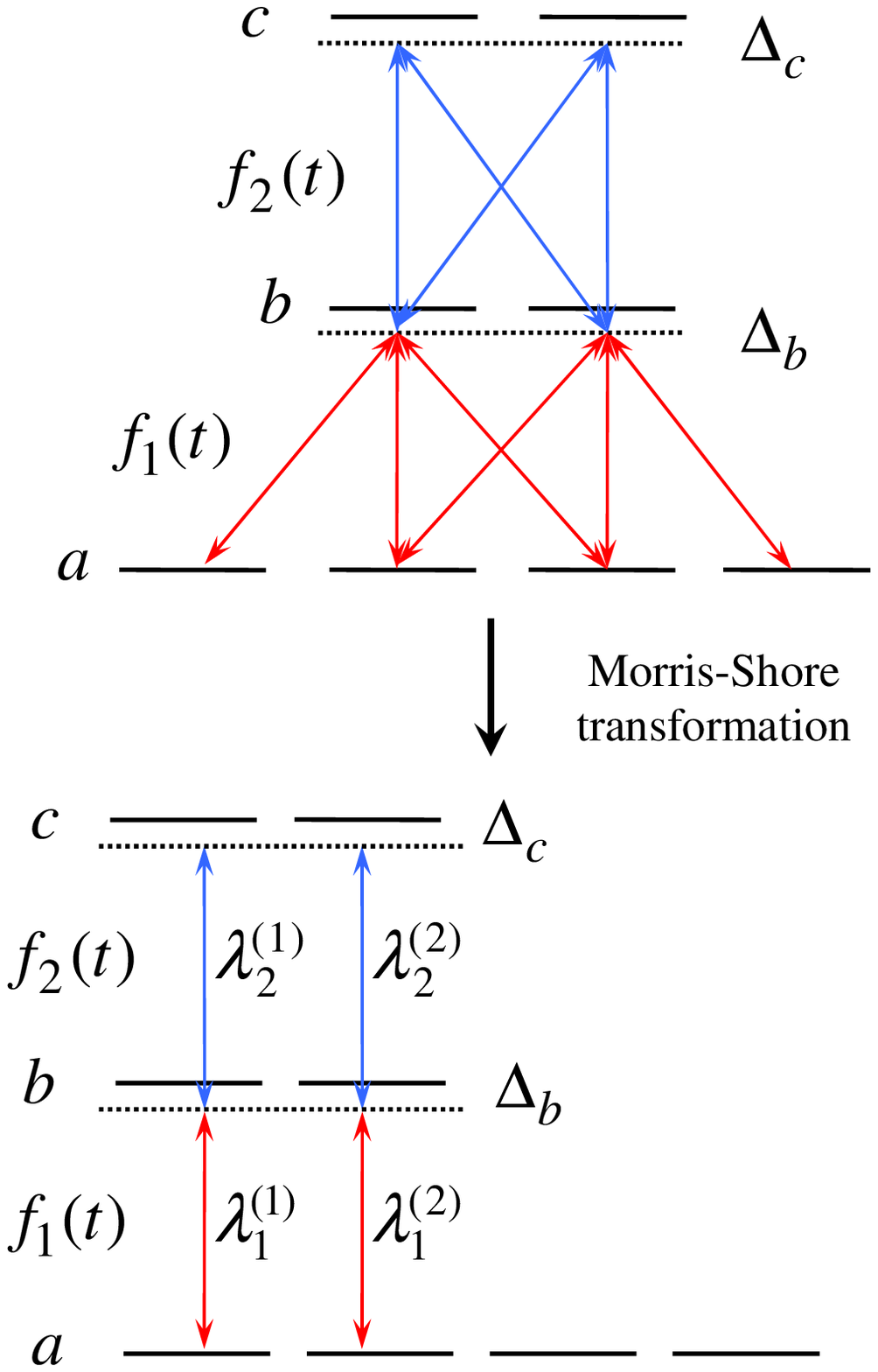}
\caption{(Color online) The MS transformation for the three-level ladder $%
J=3/2\leftrightarrow J=1/2\leftrightarrow J=1/2$: the MS decomposition
produces two independent nondegenerate three-state systems and two uncoupled
(dark) states in the $J=3/2$ set.}
\label{Fig-JJJ}
\end{figure}

With either of these choices (i) or (ii) for the fields it is possible to
reduce the original linkage pattern to a pair of three-state ladders and two
uncoupled dark states, as shown in Fig. \ref{Fig-JJJ}.

One special example for conditions (\ref{conditions.example}) is when the
left- and right-polarized fields for the lower transition have the same
intensity and the same phase ($r_{1}=l_{1}$), and the linearly-polarized
field is shifted in phase by $\pi /2$ with respect to them; then $%
l_{1}p_{1}^{\ast }+r_{1}^{\ast }p_{1}=0$. No restrictions are imposed on the
couplings of the upper transition in this case.

In another simple example, the left- and right-polarized fields for the
lower transition have the same intensity ($\left\vert r_{1}\right\vert
=\left\vert l_{1}\right\vert $) and the same applies for the upper
transition ($\left\vert r_{2}\right\vert =\left\vert l_{2}\right\vert $),
and all interactions are real.

%%%%%%%%%%%%%%%%%%%%%%%%%%%%%%%%%%%%
%%%%%%%%%%%%%%%%%%%%%%%%%%%%%%%%%%

\subsubsection{The MS\ picture}

If the commutation relation (\ref{commutator}) is satisfied then the
Hamiltonian in the MS basis reads%
\begin{equation}
\mathsf{H}^{MS}=\left[
\begin{array}{cccccccc}
0 & 0 & 0 & 0 & 0 & 0 & 0 & 0 \\
0 & 0 & 0 & 0 & 0 & 0 & 0 & 0 \\
0 & 0 & 0 & 0 & \lambda _{1}^{(1)} & 0 & 0 & 0 \\
0 & 0 & 0 & 0 & 0 & \lambda _{1}^{(2)} & 0 & 0 \\
0 & 0 & \lambda _{1}^{(1)} & 0 & \Delta _{b} & 0 & \lambda _{2}^{(1)} & 0 \\
0 & 0 & 0 & \lambda _{1}^{(2)} & 0 & \Delta _{b} & 0 & \lambda _{2}^{(2)} \\
0 & 0 & 0 & 0 & \lambda _{2}^{(1)} & 0 & \Delta _{c} & 0 \\
0 & 0 & 0 & 0 & 0 & \lambda _{2}^{(2)} & 0 & \Delta _{c}%
\end{array}%
\right] .  \label{H-example}
\end{equation}%
By rearranging the states, it is readily seen that there are two independent
three-state systems driven by Hamiltonians of the form (\ref{H3}) ($n=1,2$),
with MS couplings $\lambda _{1}^{(n)}$ and $\lambda _{2}^{(n)}$ given by
Eqs. (\ref{MS couplings}).

%-----------------------------------------

\section{Extension to $N$ levels\label{Sec-Nlevels}}

%-----------------------------------------

The results for the three-level MS transformation are readily extended to $N$
degenerate levels. For each transition $n$ ($n=1,2,\ldots ,N-1$), described
by an interaction matrix $\mathsf{V}_{n}$, time envelope $f_{n}(t)$ of all
fields in this transition, and common detuning $\Delta _{n}$, we form the
matrices $\mathsf{V}_{n}\mathsf{V}_{n}^{\dag }$. The $N$-level MS\
transformation exists if
\begin{equation}
\left[ \mathsf{V}_{n-1}^{\dag }\mathsf{V}_{n-1},\mathsf{V}_{n}\mathsf{V}%
_{n}^{\dag }\right] =\mathsf{O}\text{\textsf{\ (}}n=2,3,...,N-1\text{\textsf{%
).}}  \label{commutation}
\end{equation}%
The relations (\ref{commutation}) imply that the interactions for any two
adjacent transitions $n$ and $n+1$ must be such that, after the MS
transformation, the resulting MS states of the common level of these two
transitions are the same for the lower and upper transitions; mathematically
this is ensured by the commutation of $\mathsf{W}_{n}$ and $\mathsf{W}_{n+1}$%
.

If conditions (\ref{commutation}) are satisfied, the MS transformation will
produce sets of independent nondegenerate $N$-state systems, $(N-1)$-state
systems, and so on, and a number of uncoupled states, depending on the
particular system.

%-----------------------------------------

\section{Conclusions}

%-----------------------------------------

In this paper, we have presented an extension of the MS transformation to
three and more degenerate levels. For two degenerate sets of states the MS
transformation always exists, as long as the couplings have the same time
dependence and the same detunings. For three sets of states, the MS
transformation may or may not exist, depending of the commutation of
products of interaction matrices. When applicable, the MS transformation
reduces the coupled multistate dynamics into a set of independent
three-state systems, a set of independent two-state systems, and a number of
uncoupled (dark) states. The number of states in each set depends on the
degeneracy of the three initial sets of states. These results readily extend
to $N$ degenerate levels, with similar commutation conditions on the
interactions.

It is important that each set of states have the same RWA detuning, but this
may differ from set to set. The couplings between the first and second set
must have the same time dependence, and the same condition must apply
between the second and third set; the two time dependences, however, may be
different, as in STIRAP \cite{STIRAP}. %cccccccc
This condition extends to $N$ degenerate levels as well.

It is also important that any time dependence of the interactions is
factored out of the commutation condition (\ref{commutation}) and therefore,
for instance, different time dependence of the transition $a-b$ with respect
to the transition $b-c$ is possible. However, within each degenerate
transition ($a-b$, $b-c$, $\ldots $) the time dependence must be the same;
otherwise couplings appear between the MS states in each manifold, which
create linkages between the independent MS subsystems and the MS
decomposition does not occur.

If all detunings are zero (i.e. if all fields are on exact resonance with
the respective transition frequency) and all fields share the same time
dependence, the degenerate $N$-level system is reducible to a degenerate
two-level system; then the original MS transformation decouples the
interaction dynamics into a set of independent nondegenerate two-state
systems and a set of dark states.

In the interesting special case of a three-level ladder with a single
nondegenerate intermediate state, the MS transformation always exists, with
no restrictions on the couplings (apart from the identical time dependence).
The MS transformation produces a single linked chain, together with
additional uncoupled states.

It is significant that the MS transformation, in producing a simplification
of the original linkage pattern of interactions, introduces coherent
superpositions of the original basis states. In this MS basis the dynamics
appears very simple, and one can evaluate the conditions for producing
complete population transfer, for example. Such transitions correspond, in
the original basis, to transitions between coherent superposition states.
When the initial state is nondegenerate then only the intermediate and final
states of the transitions involve superposition states. Under such
circumstances one can design pulse sequences that produce a specified
superposition.

%-----------------------------------------

\acknowledgments

This work is supported by the EU Marie Curie ToK project CAMEL, the EU Marie
Curie RTN project EMALI, the Max-Planck Forschungspreis 2003, the Deutsche
Forschungsgemeinschaft, and the Alexander von Humboldt Foundation.

%-----------------------------------------


\begin{thebibliography}{99}
\bibitem{All75} L. Allen and J. H. Eberly, \emph{Optical Resonance and
Two-Level Atoms}, (Wiley, New York, 1975).

\bibitem{Sho90} B. W. Shore, \emph{The Theory of Coherent Atomic Excitation}%
, (Wiley, N.Y., 1990).

\bibitem{Morris83} J. R. Morris and B. W. Shore, Phys. Rev. A \textbf{27},
906 (1983).

\bibitem{MS-analytic} N. V. Vitanov, Z. Kis, and B. W. Shore, Phys. Rev. A
\textbf{68}, 063414 (2003); E. S. Kyoseva and N. V. Vitanov, Phys. Rev. A
\textbf{73}, 023420 (2006).

\bibitem{MS-CPT} Z. Kis, A. Karpati, B. W. Shore, and N. V. Vitanov, Phys.
Rev. A \textbf{70}, 053405 (2004).

\bibitem{MS-sup} Z. Kis, N. V. Vitanov, A. Karpati, C. Barthel, and K.
Bergmann, Phys. Rev. A \textbf{72}, 033403 (2005).

\bibitem{MS-tomo} N. V. Vitanov, B. W. Shore, R. G. Unanyan, and K.
Bergmann, Opt. Commun. \textbf{179}, 73 (2000); N. V. Vitanov, J. Phys. B
\textbf{33}, 2333 (2000); P. A. Ivanov and N. V. Vitanov, Opt. Commun.,
accepted (2006).

\bibitem{Bia77} Z. Bialynicka-Birula, I. Bialynicki-Birula, J. H. Eberly and
B. W. Shore, Phys. Rev. A \textbf{16}, 2048 (1977).

\bibitem{Ebe77a} J. H. Eberly, B. W. Shore, Z. Bialynicka-Birula and I.
Bialynicki-Birula, Phys. Rev. A \textbf{16}, 2038 (1977).

\bibitem{Sho81b} B. W. Shore and M. A. Johnson, Phys. Rev. A \textbf{23},
1608 (1981).

\bibitem{STIRAP} K. Bergmann, H. Theuer and B.W. Shore, Rev. Mod. Phys.
\textbf{70}, 1003 (1998); N.V. Vitanov, M. Fleischhauer, B.W. Shore and K.
Bergmann, Adv. At. Mol. Opt. Phys. \textbf{46}, 55 (2001); N.V. Vitanov, T.
Halfmann, B.W. Shore, and K. Bergmann, Ann. Rev. Phys. Chem. \textbf{52},
763 (2001).

\bibitem{Gantmacher} F. R. Gantmacher, \emph{Matrix Theory} (Springer,
Berlin, 1986).

\bibitem{Zar88} R. N. Zare, \emph{Angular Momentum: Understanding Spatial
Aspects in Chemistry and Physics}, (Wiley, N.Y., 1988).
\end{thebibliography}
\end{document}